\documentclass{cernrep} 
\usepackage{texnames}
\usepackage[T1]{fontenc}
\usepackage[bookmarks, colorlinks=true, linktoc=page, pdftex, linkcolor=red, citecolor=red, urlcolor=red]{hyperref}
\usepackage{tikz}
\usetikzlibrary{shapes.geometric, arrows}
\tikzstyle{startstop} = [rectangle, rounded corners, minimum width=3cm, minimum height=1cm,text centered, draw=black, fill=red!30]
\tikzstyle{io} = [trapezium, trapezium left angle=70, trapezium right angle=110, minimum width=3cm, minimum height=1cm, text centered, draw=black, fill=blue!30]
\tikzstyle{process} = [rectangle, minimum width=3cm, minimum height=1cm, text centered, draw=black, fill=orange!30]
\tikzstyle{decision} = [diamond, minimum width=3cm, minimum height=1cm, text centered, draw=black, fill=green!30]
\tikzstyle{arrow} = [thick,->,>=stealth]
\tikzstyle{arrow2} = [thick,<->,>=stealth]
\usetikzlibrary{matrix}
\usetikzlibrary{calc}
\usetikzlibrary{decorations.pathreplacing}

\begin{document}
\title{Dynamic simulations in SixTrack}
 
\author{K.\ Sjobak, V.K.\ Berglyd\ Olsen, R.\ De~Maria, M. Fitterer, A.\ Santamar\'{i}a Garc\'{i}a, H.\ Garcia-Morales, A.~Mereghetti, J.F.\ Wagner, S.J.\ Wretborn}

\institute{CERN, Geneva, Switzerland}

\maketitle 

\begin{abstract}
The DYNK module allows element settings in SixTrack to be changed on a turn-by-turn basis.
This document contains a technical description of the DYNK module in SixTrack.
It is mainly intended for a developer or advanced user who wants to modify the DYNK module, for example by adding more functions that can be used to calculate new element settings, or to add support for new elements that can be used with DYNK.
\end{abstract}

 
\section{Introduction}

The goal of the DYNK module in SixTrack~\cite{manual,SixPub,SixTrack_samereport,SixTrack_IPAC17} is to make it possible to change element settings on a turn-by-turn basis.
This feature was first implemented for simulating the action of scrapers in the SPS~\cite{activeCoupling}, and then re-implemented in a more general form as described in~\cite{DYNK_IPAC}.
After this re-implementation, the module has been and is being used for simulations of crab cavity failures~\cite{IPAC:CRAB:limits,IPAC:CRAB:Timescale,IPAC:CRAB:MPP,IPAC:CRAB:LLRF}, asynchronous beam dumps~\cite{Roderik:MKD}, off-momentum collimation studies through RF detuning~\cite{hector-offmomentum-here,hector-offmomentum-accnote}, studies of observed particle losses at the start of the energy ramp in the LHC~\cite{energyramping-note,energyramping-presentation}, and electron lens intensity modulation~\cite{IPAC:elens}.

The use of the module is described in the SixTrack user manual~\cite{user_manual}.
For the most up-to-date version of the user manual, please see the source distribution of SixTrack~\cite{github}.
This document describes the technical ``inner workings'' of the DYNK module, and is intended for those who want to extend the module with new mathematical functions or to handle new types of elements.

DYNK is designed to be extensible, and has already been extended several times by several people.
It is the hope of the authors that this document will be helpful for those who wish to do this in the future.
Note that it is strongly recommended to read about the usage of DYNK in the most recent version of the SixTrack user manual before reading this document.

\section{Overview}

The main code of DYNK is located in the Fortran module \texttt{dynk} found in the \texttt{dynk.s90} file.
The \texttt{.s90} file format is free form Fortran (\texttt{.f90}) which will be pre-processed by ASTUCE\footnote{%
ASTUCE is a source code preprocessor that is used for SixTrack.
It assembles the Fortran source files from one or more decks, typically containing a group of related subroutines; and blocks, small pieces of codes that are in-lined one or more places in one or more decks, typically used for defining Fortran COMMON blocks and kicks.
Furthermore, the ASTUCE ``pragmas'' can include conditional statements, used to include or exclude parts of the source code based on compilation flags.}.
All the data associated with DYNK is contained within the \texttt{dynk} module as saved arrays and variables. 

Additionally, the subroutine \texttt{initialize\_element} was added to the \texttt{daten} block.
It is responsible for general initialization of \texttt{SINGLE ELEMENT}s at the end of the \texttt{daten} subroutine, whether DYNK is in use or not.
It is also used to re-initialize such elements if they are later changed by DYNK.

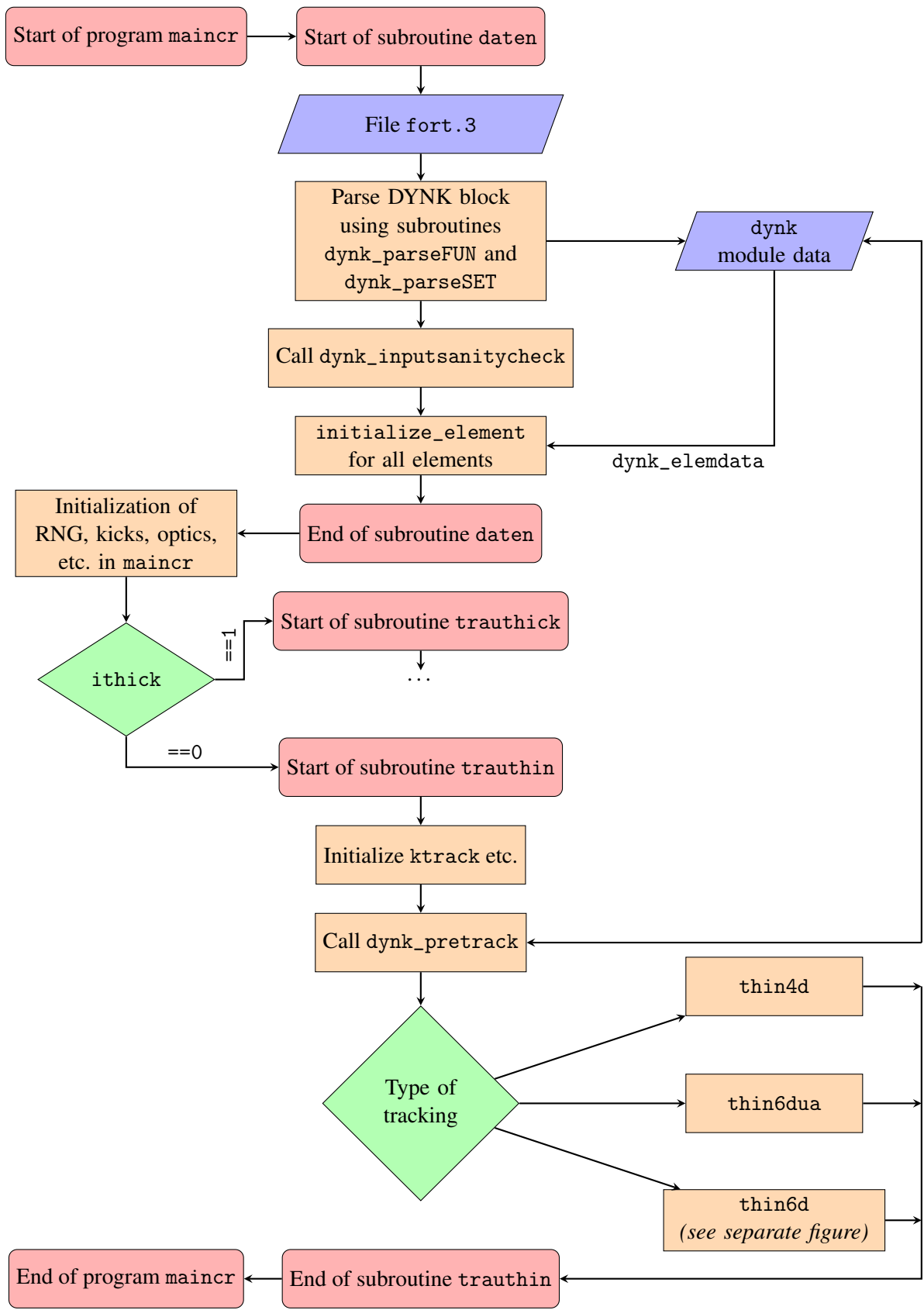
\begin{figure}[p]
    \centering
    \begin{tikzpicture}[node distance=2cm]
    \node (maincr_start) [startstop] {Start of program \texttt{maincr}};
    
    \node (daten) [startstop,right of=maincr_start, xshift=3cm] {Start of subroutine \texttt{daten}};
    \node (fort3) [io,below of=daten,yshift=0.5cm] {File \texttt{fort.3}};
    \node (dynk_parse) [process, below of=fort3, text width=4cm] {Parse DYNK block using subroutines \texttt{dynk\_parseFUN} and \texttt{dynk\_parseSET}};
    \node (comdynk) [io, right of=dynk_parse, xshift=4cm, text width=2.3cm] {\texttt{dynk} module data};
    \node (sanitycheck) [process, below of=dynk_parse] {Call \texttt{dynk\_inputsanitycheck}};
    \node (init_elem_1) [process, below of = sanitycheck, text width=4cm,yshift=.5cm] {\texttt{initialize\_element} for all elements};
    \node (daten_end) [startstop,below of=init_elem_1,yshift=.5cm] {End of subroutine \texttt{daten}};
    
    \draw[arrow] (maincr_start) |- (daten);
    \draw[arrow] (daten) -- (fort3);
    \draw[arrow] (fort3) -- (dynk_parse);
    \draw[arrow] (dynk_parse) -- (comdynk);
    \draw[arrow] (dynk_parse) -- (sanitycheck);
    \draw[arrow] (sanitycheck) -- (init_elem_1);
    \draw[arrow] (init_elem_1) -- (daten_end);
    \draw[arrow] (comdynk) |- node[anchor=north, text width=2cm,xshift=-1.75cm] {\texttt{dynk\_elemdata}} (init_elem_1);
    
    \node (maincr_init) [process, left of=daten_end, xshift=-3cm,yshift=0cm, text width=3.5cm] {Initialization of RNG, kicks, optics, etc.\ in \texttt{maincr}};
    \node (ithick) [decision, below of = maincr_init,yshift=-0.5cm] {\texttt{ithick}};
    
    \draw[arrow] (daten_end) -- (maincr_init);
    \draw[arrow] (maincr_init) -- (ithick);
    
    \node (thick) [startstop, right of=ithick, xshift=3cm,yshift=1cm] {Start of subroutine \texttt{trauthick}};        
    \node(thick2) [below of = thick,yshift=1cm] {\ldots};
    \draw[arrow] (thick) -- (thick2);
    \draw[arrow] (ithick) -- +(2cm,0cm) node[yshift=0.6cm,rotate=90,anchor=south]{\texttt{==1}} |- (thick);
    
    \node (thin) [startstop, right of=ithick,yshift=-1.5cm, xshift=3cm] {Start of subroutine \texttt{trauthin}};
    \draw[arrow] (ithick) |- node[anchor=south, xshift=1cm] {\texttt{==0}} (thin);
    
    \node (init_thin) [process, below of = thin, yshift=0.5cm] {Initialize \texttt{ktrack} etc.};
    \node (dynk_pretrack) [process, below of = init_thin, yshift=0.5cm] {Call \texttt{dynk\_pretrack}};
    \node (dims) [decision, below of = dynk_pretrack, text width = 2cm, yshift=-0.75cm] {Type of tracking};
    
    \draw[arrow] (thin) -- (init_thin);
    \draw[arrow] (init_thin) -- (dynk_pretrack);
    \draw[arrow] (dynk_pretrack) -- (dims);
    
    \draw[arrow] (dynk_pretrack)+(8.5cm,0) -- (dynk_pretrack);
    \draw[arrow] (dynk_pretrack)+(8.5cm,0) |- (comdynk);
    
    \node[process, right of = dims, xshift=4cm, yshift=2cm] (thin4d) {\texttt{thin4d}};
    \node[process, right of = dims, xshift=4cm, yshift=0cm] (thin6dua) {\texttt{thin6dua}};
    \node[process, right of = dims, xshift=4cm, yshift=-2cm, text width=3.5cm] (thin6d) {\texttt{thin6d}\\ \textit{(see separate figure)}};
    
    \node (trauthin_stop) [startstop, below of=dims, yshift=-1cm] {End of subroutine \texttt{trauthin}};
    
    \draw[arrow] (dims) -- (thin4d);
    \draw[arrow] (dims) -- (thin6dua);
    \draw[arrow] (dims) -- (thin6d);
    
    \draw[arrow] (thin4d) -- +(2.5cm,0); 
    \draw[arrow] (thin6dua) -- +(2.5cm,0);
    \draw[arrow] (thin6d) -- +(2.5cm,0);
    \draw[arrow] (thin4d)+(2.5cm,0) |- (trauthin_stop);
    
    \node (maincr_stop) [startstop, left of=trauthin_stop, xshift=-3cm] {End of program \texttt{maincr}};
    
    \draw[arrow] (trauthin_stop) -- (maincr_stop);
\end{tikzpicture}
    \caption{Program flow in SixTrack with DYNK. The details of \texttt{thin6d} is shown in Figure~\ref{fig:thin6d}. The beginning/end of processes are in red blocks, orange blocks are actions of interest, blue blocks are I/O, and green blocks are decisions.}
    \label{fig:flow}
\end{figure}

\begin{figure}[p]
    \centering
    \begin{tikzpicture}[node distance = 1.5cm]
    \node[startstop, text width=3.5cm] (select_start) {Have selected \texttt{thin6d} (still in \texttt{trauthin})};
    \node[decision, below of = select_start,yshift=-1.0cm, text width = 1.75cm] (collimat1) {Collimation version?};
    \node[process, below of = collimat1, yshift=-1.25cm] (collimat2) {Initialize collimation};
    \node[startstop, below of =collimat2] (collimat3) {Begin loop over samples};
    \node[process, below of = collimat3, text width=3cm] (collimat4) {Load particles in tracking arrays};
    
    \draw[arrow] (select_start) -- (collimat1);
    \draw[arrow] (collimat1) -- node[xshift=0.5cm]{yes} (collimat2);
    \draw[arrow] (collimat2) -- (collimat3);
    \draw[arrow] (collimat3) -- (collimat4);
    
    \node[startstop, right of = collimat1,xshift=3.5cm] (thin6d_start) {Beginning of subroutine \texttt{thin6d}};
    \draw[arrow] (collimat1) -- node[xshift=0cm,yshift=0.25cm]{no} (thin6d_start);
    \draw[arrow] (collimat4) -| (thin6d_start.192) ;
    \node[process, below of=thin6d_start] (thin6d_init) {Initialize \texttt{thin6d}};
    
    \node[startstop, below of = thin6d_start] (thin6d_turns1) {Begin loop over turns};
    
    \node[startstop, below of = thin6d_turns1, xshift=3cm, text width=4cm] (dynkapply_start) {Beginning of\\ subroutine \texttt{dynk\_apply}};
    
    \draw[arrow] (thin6d_start) -- (thin6d_init);
    \draw[arrow] (thin6d_init) |- (dynkapply_start);
    
    \node[process,below of = dynkapply_start,text width=4.2cm,yshift=-0.25cm] (dynkapply_init) {Initialize \texttt{dynk\_apply},\\ reset DYNK and elements if turn=1 (collimation) };
    \node[process, below of = dynkapply_init, text width = 5cm,yshift=-1cm] (dynkapply_change) {
        Change settings:
        \vspace{-0.25cm}
        \begin{enumerate}
        \item \texttt{dynk\_computeFUN}
        \item \texttt{dynk\_setvalue}
        \item \texttt{(initialize\_element)}
        \end{enumerate}
        \vspace{-0.5cm}
    };
    \node[process, below of = dynkapply_change, text width = 6cm,yshift=-0.75cm] (dynkapply_check) {
        Confirm settings (\texttt{dynk\_getvalue}),\\
        write to file \texttt{dynksets.dat}
    };
    \node[startstop,below of = dynkapply_check] (dynkapply_end) {End of \texttt{dynk\_apply}};
    \node[io, right of = dynkapply_check,xshift=3cm, text width = 2.75cm, minimum width=3cm,rotate=90] (dynksetsdat) {File\\ \texttt{dynksets.dat}};
    
    \node[io, right of = dynkapply_init,xshift=3cm, text width=2.75cm, rotate=90] (comdynk) {\texttt{dynk} module data};
    
    \draw[arrow] (dynkapply_start)--(dynkapply_init);
    \draw[arrow] (dynkapply_init)--(dynkapply_change);
    \draw[arrow] (dynkapply_change)--(dynkapply_check);
    \draw[arrow] (dynkapply_check)--(dynksetsdat);
    \draw[arrow] (dynkapply_check)--(dynkapply_end);
    
    \draw[arrow2] (dynkapply_init) -- (comdynk);
    \draw[arrow2] (dynkapply_change) -| (comdynk);
    
    \node[process, below of = dynkapply_end, xshift=-3cm,text width = 4cm] (kicks) {Loop over elements,\\ apply kicks};
    
    \node[startstop, below of = kicks] (thin6d_turns2) {End loop over turns};
    
    \draw[arrow] (dynkapply_end) -| (kicks);
    \draw[arrow] (kicks) -- (thin6d_turns2);
    
    \draw[arrow] (thin6d_turns2.180) -- ([xshift=-2cm] thin6d_turns2.180) node[yshift=3cm,xshift=-0.25cm,rotate=90]{If more turns} |- ([yshift=-5cm] thin6d_turns1.200) -| (thin6d_turns1.200);
    
    \node[startstop, below of = thin6d_turns2] (thin6d_end) {End of subroutine \texttt{thin6d}};
    \node[startstop, left of = thin6d_end,xshift=-3.5cm, text width = 4cm] (samples_end) {If collimation version:\\ End loop over samples};
    \draw[arrow] (thin6d_turns2)--(thin6d_end);
    \draw[arrow] (thin6d_end)--(samples_end);
    
    \node[startstop, below of=samples_end] (thin6d_end) {End of subroutine \texttt{trauthin}};
    \draw[arrow] (samples_end)--(thin6d_end);
    \draw[arrow] (samples_end.180) -- ([xshift=-1cm]samples_end.180) node[yshift=4cm,xshift=-0.25cm,rotate=90]{If more samples} |- (collimat3.180);
\end{tikzpicture}
    \caption{Details of program flow when running subroutine \texttt{thin6d}; see Figure~\ref{fig:flow} for the rest of the program. The beginning/end of processes are in red blocks, orange blocks are actions of interest, blue blocks are I/O, and green blocks are decisions.}
    \label{fig:thin6d}
\end{figure}
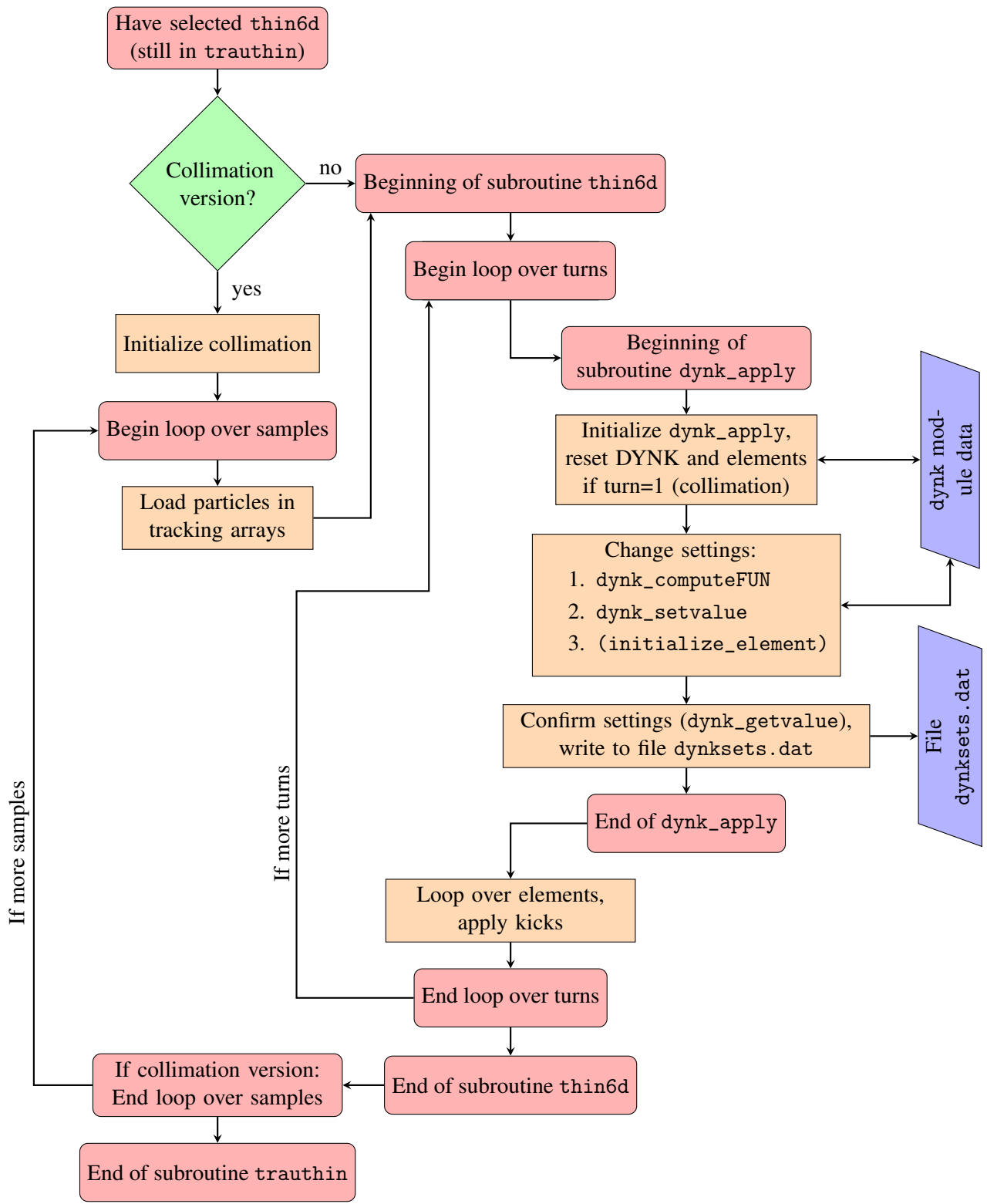

The general program flow is illustrated in Figures~\ref{fig:flow} and~\ref{fig:thin6d}.
These show how DYNK is initialized, and how it is called.
The structure is only shown for 6D thin tracking, however the pattern is the same for the other options.

Note that all DYNK-related global variables have names postfixed with \texttt{\_dynk}, while all functions and subroutines related to DYNK bear the prefix \texttt{dynk\_}.
Since the subroutine \texttt{initialize\_element} is also used outside of DYNK, it does not contain this pre- or post-fix in its name.
Furthermore, while this document is intended to give an overview of how DYNK functions, detailed descriptions of each subroutine, code stub, and variable are given in the code as comments.

\section{Data structure}
\label{sec:data}

The configuration of DYNK is stored in the \texttt{dynk} module.
These variables and arrays are separated in 3 main categories, as described in the following subsections.

\subsection{Overall configuration}
Whether DYNK is active at all, i.e.\ if a DYNK block is present in the \texttt{fort.3} input file, is controlled by the \texttt{ldynk} boolean variable.
If this is \texttt{.FALSE.}, no DYNK functions are called in the tracking loop.
Furthermore, it is possible to enable extra debugging output and run-time checks (\texttt{ldynkdebug}), and to disable writing of the \texttt{dynksets.dat} output file (\texttt{ldynkfiledisable}).

\subsection{Functions}
\label{sec:data:FUN}

\begin{figure}[htb]
    \centering
    \begin{tikzpicture}[node distance=2cm]
	\matrix[matrix of nodes,column 1/.style={anchor=base west}] (funcs) {
    	Statement in \texttt{fort.3}    &       & Row & Name & Type & Data 1 & Data 2 & Data 3 \\
  		\hline
    	|[red]| \texttt{FUN f1 FILE myfile.txt} &  $\to$ & |[red]| 1 & |[red]| 1 & |[red]| 1 & |[red]|     2 & |[red]|     1 & |[red]|     5 \\
        |[blue]| \texttt{FUN f2 LIN 2.5 3.14}   & $\to$ & |[blue]| 2 & |[blue]| 3 & |[blue]| 42 & |[blue]| 6 & |[blue]| -1 & |[blue]| -1 \\
        |[brown]| \texttt{FUN f3 ADD f1 f2} & $\to$ & |[brown]| 3   & |[brown]| 4 & |[brown]| 20 & |[brown]| 1 & |[brown]| 2 & |[brown]| -1 \\
    };
    \draw({$(funcs-1-3)!.5!(funcs-1-4)$} |- funcs.north) -- ({$(funcs-4-3)!.5!(funcs-4-4)$} |- funcs.south);
    \node[anchor=west, yshift=0.25cm] at (funcs.north west) {\texttt{funcs\_dynk:}};
    \node[anchor=west, yshift=-0.25cm] at (funcs.south west) {\texttt{nfuncs\_dynk = 3 $\le$ maxfuncs\_dynk}};
    
    \matrix[matrix of nodes, below of=funcs, yshift=-2.5cm,xshift=-3.75cm] (cexpr) {
    	Row & Data (string) \\
        \hline
        |[red]| 1   & |[red]| f1 \\
        |[red]| 2   & |[red]| myfile.txt \\
        |[blue]| 3   & |[blue]| f2 \\
        |[brown]|4   & |[brown]| f3 \\
    };
    \draw({$(cexpr-1-1)!.3!(cexpr-1-2)$} |- cexpr.north) -- ({$(cexpr-5-1)!.3!(cexpr-5-2)$} |- cexpr.south);
    \node[anchor=west, yshift=0.25cm] at (cexpr.north west) {\texttt{cexpr\_dynk:}};
    \node[anchor=west, yshift=-0.25cm] at (cexpr.south west) {\texttt{ncexpr\_dynk = 4 $\le$ maxdata\_dynk}};
    
    \matrix[matrix of nodes, below of=funcs, yshift=-2.5cm,xshift=3cm] (fexpr) {
    	Row & Data (real num.) \\
        \hline
        |[red]| 1 & |[red]| 23.5 \\
        |[red]| 2 & |[red]| 27.8 \\
        |[red]| 3 & |[red]| 20.0 \\
        |[red]| 4 & |[red]| 15.2 \\
        |[red]| 5 & |[red]| 30.0 \\
        |[blue]| 6 & |[blue]| 2.5  \\
        |[blue]| 7 & |[blue]| 3.14 \\
    };
    \draw({$(fexpr-1-1)!.3!(fexpr-1-2)$} |- fexpr.north) -- ({$(fexpr-5-1)!.3!(fexpr-5-2)$} |- fexpr.south);
    \node[anchor=west, yshift=0.25cm] at (fexpr.north west) {\texttt{fexpr\_dynk:}};
	\node[anchor=west, yshift=-0.25cm] at (fexpr.south west) {\texttt{nfexpr\_dynk = 7 $\le$ maxdata\_dynk}};
    
    \draw[brown,->] (funcs-4-6) -- (funcs-2-3);
    \draw[brown,->] (funcs-4-7) -- (funcs-3-3);
    
    \draw[red,->] (funcs-2-4) -- (cexpr-2-1);
    \draw[blue,->] (funcs-3-4) -- (cexpr-4-1);
    \draw[brown,->] (funcs-4-4) -- (cexpr-5-1);
    
    \draw[red,->] (funcs-2-6) -- (cexpr-3-1);
    \draw[red,->] (funcs-2-7) -- (fexpr-2-1);
    \draw[blue,->] (funcs-3-6) -- (fexpr-7-1);
    
    \draw[decorate,decoration={brace,amplitude=10pt,mirror},red] (fexpr-6-2.south east) --  (fexpr-2-2.north east);
    \draw[red,->] (funcs-2-8) -- ($ (fexpr-4-2.east) + (10pt,0) $);
    
\end{tikzpicture}
    \caption{Illustration of how three DYNK functions are stored in memory. The colours are used to separate the three functions used in the example ({\color{red} f1}, {\color{blue} f2}, {\color{brown} f3}). The function {\color{blue} f1} is of type FILE, where the column Data1 points to where in \texttt{fexpr\_dynk} the data begins, and Data2 indicates how much data there is. The function {\color{blue}f2} is of type LIN, where Data1 points to where in \texttt{fexpr\_dynk} the two parameters are stored. The function {\color{brown}f3} is of type ADD, where Data1 and Data2 indicates which functions should be added by listing their row numbers in \texttt{funcs\_dynk}.}
    \label{fig:mem-FUN}
\end{figure}
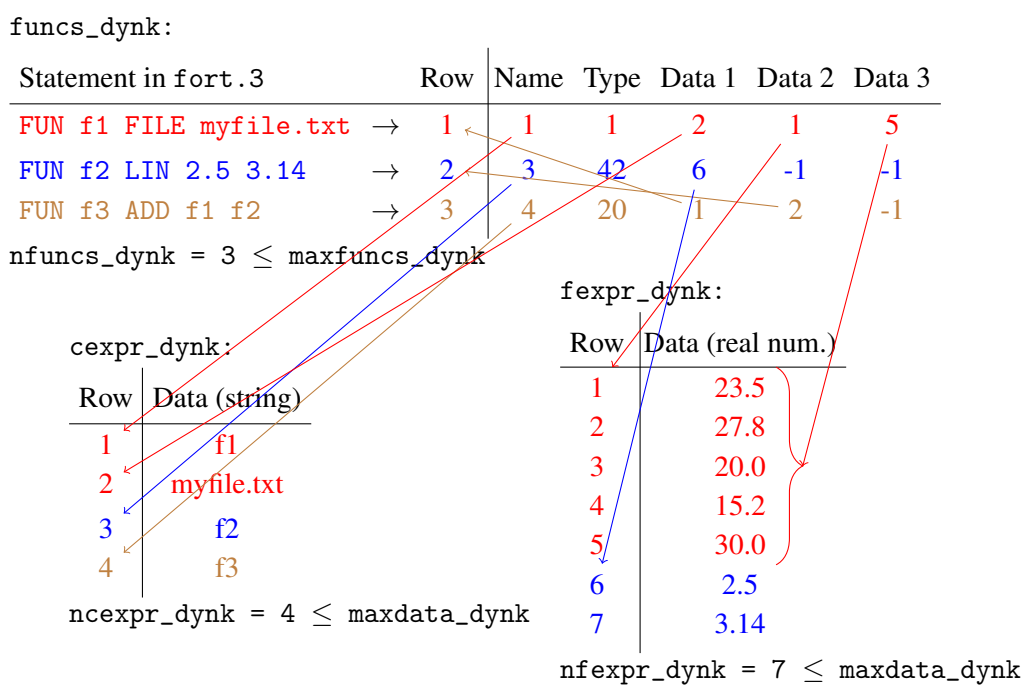

At the core of DYNK is the possibility to compute function values.
This can for instance be used to change the strength of elements as a function of the turn number.
The details of how this works is described in Section~\ref{sec:FUN}.
As illustrated in Figure~\ref{fig:mem-FUN}, the configuration for these functions is described in the table \texttt{funcs\_dynk}, with one row per function (\texttt{FUN} statements in \texttt{fort.3}).
Each of these rows consist of five integers, for which the first two have the same meaning for all function types.
The meaning of the last three columns vary.
Here, the first column always points to the location in the \texttt{cexpr\_dynk} array where the (user-defined) name of the function is stored, while the second is an index which indicates the type of the function, and thus the interpretation of the following three columns.
The number of function ``slots'' currently in use is defined by the variable \texttt{nfuncs\_dynk} such that the last row that is currently in use is row number \texttt{nfuncs\_dynk}\footnote{Note the Fortran convention of counting from 1.}.
The size of the table, and thus the maximum amount of slots, is given by the constant parameter \texttt{maxfuncs\_dynk}.

Since many functions require the storage of more than three integers, DYNK has an internal memory allocation mechanism supporting the three main data types: integers, double precision real numbers, and strings.
This is implemented as three large arrays \texttt{iexpr\_dynk}, \texttt{fexpr\_dynk}, and \texttt{cexpr\_dynk}.
Each string has a maximum length \texttt{maxstrlen\_dynk}, and the stored strings are generally expected to be terminated and padded with binary zeros.
Similar to \texttt{nfuncs\_dynk}, each of these arrays have an associated integer that keeps track of how many elements are currently in use.
These usage counters are called \texttt{niexpr\_dynk}, \texttt{nfexpr\_dynk}, and \texttt{ncexpr\_dynk}, and should point to the last valid index in their corresponding data array.

It is the developer's responsibility to make sure that a new function (1) does not interfere with the data storage for other functions, and (2) does not exceed the available storage for this data type.
For the first point, this means that the function must not modify the data stored by other functions, and it must correctly update the usage counters so that the next function to be defined can safely allocate more memory.
In order to help with the second point, the subroutine \texttt{dynk\_checkspace} is provided, which dynamically expands the arrays when needed.
The expansion is normally done in chunks of 500 numbers or 200 strings.
If more than this is requested, the allocation is exactly large enough to fit the requested amount.

\subsection{Element settings}
\label{sec:data:SET}

\begin{figure}[htb]
    \centering
    \begin{tikzpicture}[node distance=2cm]
	\matrix[matrix of nodes, nodes={font=\small},column 1/.style={anchor=base west}] (sets) {
    	Statement in \texttt{fort.3}                &       & Row & FUN idx. & First & Last & Shift & Element & Attribute\\
    	\hline
    	\texttt{SET magnet1 average\_ms f3 1 5 0}   & $\to$ &   1 & 3 & 1 & 5  &  0 & magnet1 & average\_ms \\
    	\texttt{SET magnet1 average\_ms f3 6 -1 -5} & $\to$ &   2 & 3 & 6 & -1 & -5 & magnet1 & average\_ms \\
    	\texttt{SET crabcc1 voltage f2 3 10 -3}     & $\to$ &   3 & 2 & 3 & 10 & -3 & crabcc1 & voltage \\
     };
    \draw({$(sets-2-3)!.35!(sets-2-4)$} |- sets.north) -- ({$(sets-4-3)!.35!(sets-4-4)$} |- sets.south);
    \draw({$(sets-2-7)!.4!(sets-2-8)$} |- sets.north) -- ({$(sets-4-7)!.4!(sets-4-8)$} |- sets.south);
    \node[anchor=west, yshift=-0.25cm] at (sets.south west) {\texttt{nsets\_dynk = 3 $\le$ maxsets\_dynk}};
    
    \draw[decorate,decoration={brace,amplitude=10pt},yshift=100pt] (sets-1-4.north west) -- node[yshift=15pt] {\texttt{sets\_dynk:}} (sets-1-7.north east);
    \draw[decorate,decoration={brace,amplitude=10pt},yshift=100pt] (sets-1-8.north west) -- node[yshift=15pt] {\texttt{csets\_dynk:}} (sets-1-9.north east);
    
    \matrix[matrix of nodes, nodes={font=\small},below of=sets, yshift=-1.5cm] (sets2) {
   		Row & Element  & Attribute \\
    	\hline
    	  1 & magnet1 & average\_ms \\
    	  2 & crabcc1 & voltage \\
     };
    \draw({$(sets2-2-1)!.35!(sets2-2-2)$} |- sets2.north) -- ({$(sets2-3-1)!.35!(sets2-3-2)$} |- sets2.south);
	\node[anchor=west, yshift=0.25cm] at (sets2.north west) {\texttt{sets\_unique\_dynk:}};
    \node[anchor=west, yshift=-0.25cm] at (sets2.south west) {\texttt{nsets\_unique\_dynk = 2 $\le$ maxsets\_dynk}};
    
\end{tikzpicture}
    \caption{Illustration of how three \texttt{SET} statements are stored in memory.}
    \label{fig:mem-SET}
\end{figure}

Similar to the functions, the element settings (\texttt{SET} statements in \texttt{fort.3}) are also stored in a few tables.
However, since all settings are defined by the same fields, the memory management is considerably simpler.
As illustrated in Figure~\ref{fig:mem-SET} the main table is \texttt{sets\_dynk}, which can store up to \texttt{maxsets\_dynk} rows of integers.
Each row of this table contains four columns.
The first column is pointing to a row in the \texttt{funcs\_dynk} array, indicating which function should be used for this element setting.
The two next columns indicate the range of turns for which the setting is active.
The last column is an offset which is applied to the current turn number before computing the value of the function.

Additionally, there is a table \texttt{csets\_dynk}, which has one row per \texttt{SET} with the same indexing as \texttt{sets\_dynk}).
Each row has two columns, both strings.
The first column is the name of the \texttt{SINGLE ELEMENT} to be changed, and the second column is the name of the attribute that should be changed.
This table is filled in the subroutine \texttt{dynk\_pretrack}.

This architecture allows multiple functions to be defined for the same element, as long as the range of turns they are used for is not overlapping.
This is verified by the \texttt{dynk\_pretrack} subroutine, which also verifies that all \texttt{SET}s refer to valid elements and attributes.
While this data structure makes it possible to apply different functions to the same element and attribute for different time periods, in many cases it is necessary to directly iterate over the element and attribute combinations that are used instead of the individual SET statements.
To avoid excessive searching when iterating over unique elements and attributes, the table \texttt{csets\_unique\_dynk} is used.
The table has the same format as \texttt{csets\_dynk}, but has no duplicate entries.
The number of rows used in this table is kept in the integer variable \texttt{nsets\_unique\_dynk}.
Furthermore, the array \texttt{fsets\_origvalue\_dynk} contains the pre-changed value of the elements using the same indexing as \texttt{csets\_unique\_dynk}.
This is necessary for collimation, where multiple samples of particles are tracked sequentially, making it necessary to be able to reset the state of the elements to how it was originally defined in \texttt{fort.2} and \texttt{fort.3}.

Finally, there are two arrays that are used to keep additional information on the level of structure elements, of which there may be several per single element.
These arrays are named \texttt{dynk\_izuIndex} and \texttt{dynk\_elemdata}.
The array \texttt{dynk\_izuIndex} is used to store the index of the random number used to set the magnet error of the element in question.
The array \texttt{dynk\_elemdata} is used to store the various values needed to initialize certain types of elements, such as accelerating RF cavities, and will most likely be necessary for handling magnetic multipoles should this be implemented in the future.

\section{Changing element settings: \texttt{dynk\_apply}}

If DYNK is active in the simulation, the subroutine \texttt{dynk\_apply} is called in the tracking loop at the beginning of every turn, as shown in Figure~\ref{fig:thin6d}.
This subroutine then loops over the defined \texttt{SET}s, checking if it is active in the current turn.
If this is the case, the current effective turn is calculated by adding the turn-shift (which is often 0) to the actual turn.
The value of the active function is then calculated using \texttt{dynk\_computeFUN} as described in Section~\ref{sec:FUN}.

Note that \texttt{dynk\_apply} has an important role in initializing DYNK \texttt{FUN}ctions where the output depends on the internal state created by previous calls.
This happens with the \texttt{FIR} and \texttt{IIR} filters, and also with the RNG-based functions \texttt{RANDG}, \texttt{RANDU}, and \texttt{RANDON}.
This is done on the first turn, before any kicks are applied.
The action taken is for example to initialize the ``seed'' that is updated at every call to the RNG function, with the initial seed read from \texttt{fort.3}.

\subsection{The subroutine \texttt{dynk\_setvalue}}
When the new value has been calculated, it is applied to the specified element and attribute using \texttt{dynk\_setvalue}.
This function modifies the main element setting arrays in the same way as \texttt{daten} does when reading the input files, doing no calculations.
In practice, this means that when changing single elements, it will write to the \texttt{ed}/\texttt{ek}/\texttt{el} and \texttt{elens\_theta\_max} arrays.
The reason for this is to make the code, and its testing, as straight forward as possible.

One exception to the ``no calculations'' rule is that when changing the reference energy \texttt{E0} using \texttt{GLOBAL-VARS}, which triggers a recalculation of the reference momentum \texttt{e0f}, relativistic gamma \texttt{gammar}, and the energy-dependent particle arrays \texttt{dpsv}/\texttt{dpsv1}/\texttt{dpd}/\texttt{dpsq}/\texttt{oidpsv}/\texttt{rvv}.

\subsection{The subroutine \texttt{initialize\_element}}
If any further calculation is needed, this is done afterwards in the subroutine \text{initialize\_element}.
This is for example used for nonlinear elements, where the actual kick strength is given by an average strength (\texttt{ed}) and a random component (\texttt{ek}).
Furthermore, a unit conversion is also applied.

Another example is crab cavities, which store the phase offset in \texttt{el} when reading the input file, which is normally used to store the element length.
This phase is therefore moved into a separate array \texttt{crabph}, which has the same shape and indexing as \texttt{el}.

Note that \texttt{initialize\_element} is not only used when changing element settings, but also used just after reading the input files in \texttt{daten}.
In this case, the second argument ``\texttt{lfirst}'' is set to \texttt{.TRUE.}.

\subsection{The subroutine \texttt{dynk\_getvalue}}
The function \texttt{dynk\_getvalue} returns the setting of an element's attribute as it was set by \texttt{dynk\_setvalue}.
This is used for output to \texttt{dynksets.dat}.
When \texttt{ldynkdebug=.TRUE.} it is also used to confirm that the setting was correctly applied.

\subsection{The virtual element \texttt{GLOBAL-VARS}}
In order to change certain global settings, such as the reference energy \texttt{E0}, the element name \texttt{GLOBAL-VARS} is used.
Because of this, it is not possible to have an actual element in the single elements list with this name when DYNK is active.

The \texttt{GLOBAL-VARS} element is also treated specially in the \texttt{dynk\_setvalue} and \texttt{dynk\_getvalue} subroutines.
Here, this ``element name'' is tested for and handled near the top of the routine, before trying to locate the normal elements.

\section{How the functions are calculated}
\label{sec:FUN}

\begin{table}
\centering
\begin{tabular}{c c|l}
Index & Name & Short description\\
\hline
\multicolumn{2}{c|}{"System" functions:}& \\
0 & \texttt{GET} & Get original value of element/attribute.\\
1 & \texttt{FILE} & Load from file (turn-by-turn).\\
2 & \texttt{FILELIN} & Load from file (interpolate non-specified turns).\\
3 & \texttt{PIPE} & Get setting from external program.\\
6 & \texttt{RANDG} & Gaussian RNG.\\
7 & \texttt{RANDU} & Uniform RNG.\\
8 & \texttt{RANDON} & Random 1 or 0 with given probability.\\
\hline
\multicolumn{2}{c|}{Filters:} & \\
10 & \texttt{FIR} & Finite Impulse Response filter.\\
11 & \texttt{IIR}  & Infinite Impulse Response filter. \\
\hline
\multicolumn{2}{c|}{Operators (2-operand):} & \\
20 & \texttt{ADD} & Add the results of two other functions.\\
21 & \texttt{SUB} & Subtract the results of two other functions.\\
22 & \texttt{MUL} & Multiply the results of two other functions.\\
23 & \texttt{DIV} & Divide the results of one function by the result of another.\\
24 & \texttt{POW} & Exponentiate the result of one function with the result of another.\\
\hline
\multicolumn{2}{c|}{Operators (1-operand):} & \\
30 & \texttt{MINUS} & Result of another function, with opposite sign. \\
31 & \texttt{SQRT} & Square root of the result from another function.\\
32 & \texttt{SIN} & Sine of the result from another function.\\
33 & \texttt{COS} & Cosine of the result from another function.\\
34 & \texttt{LOG} & Natural logarithm of the result from another function.\\
35 & \texttt{LOG10} & Base-10 logarithm of the result from another function.\\
36 & \texttt{EXP} & Natural exponential function of the result from another function.\\
\hline
\multicolumn{2}{c|}{Polynomial functions:} & \\
40 & \texttt{CONST} & Constant value.\\
41 & \texttt{TURN} & Current turn (after turn-shift).\\
42 & \texttt{LIN} & Linear function of the turn number.\\
43 & \texttt{LINSEG} & Linear function of the turn number (alternative input format).\\
44 & \texttt{QUAD} & Quadratic function of the turn number.\\
45 & \texttt{QUADSEG} & Quadratic function of the turn number (alternative input format).\\
\hline
\multicolumn{2}{c|}{Transcendental functions:} & \\
60 & \texttt{SINF} & Sine of the turn number, with specified $\omega$, $\phi$, and $A$.\\
61 & \texttt{COSF} & Cosine of the turn number, with specified $\omega$, $\phi$, and $A$.\\
62 & \texttt{COSF\_RIPP} & Cosne of the turn number (alternate ``RIPP'' format).\\
\hline
\multicolumn{2}{c|}{Specialized functions:} & \\
80 & \texttt{PELP}  & Parabolic-Exponential-Linear-Parabolic, used for energy ramping~\cite{Russenschuck:PELP}.\\
81 & \texttt{ONOFF} & On for $p_1$ turns, then off for $p_2-p_1$ turns, then repeat.
\end{tabular}
\caption{Functions defined in DYNK and their indexes. For a full description,  please see the user manual.}
\label{tab:FUN}
\end{table}

As shown in Table~\ref{tab:FUN}, many types of functions are available in DYNK.
The data used for their evaluation is stored in the \texttt{dynk} module, as described in Section~\ref{sec:data:FUN}.
Internally, each function type is given an integer index, which is used in a \texttt{select case} when evaluating the function.
Note that these indices are grouped by function type, and this grouping should be respected when adding new functions.
A detailed description of the use of each function is given in the user manual, and how they store their data is documented in comments in the \texttt{dynk\_parseFUN} code.

In most cases the DYNK \texttt{FUN}s are functions of the current turn number, including a possible turn-shift, which may be specified in the \texttt{SET} command.
Note that DYNK functions can be chained such that one function can call another one, which may call yet another, etc.
This is done by for example the \texttt{ADD} function, which calls two other functions, adds their results together, and returns the value.
When this happens, the current turn number -- including any turn-shift -- is passed on to the next function.

\subsection{Calculating the function values f(turn): \texttt{dynk\_computeFUN}}
\label{sec:FUN:compute}

The values of the functions are calculated in this Fortran function.
It is called with an index into the relevant row in the \texttt{funcs\_dynk} table, and the current (possibly shifted) turn number.
Using the row index, it then reads the type index (second column), and executes the case containing the code for that function.
The code may then read or modify any data in the \texttt{funcs\_dynk} array and the \texttt{\{c|f|i\}expr\_dynk} arrays.
Note that it is expected that functions should only touch data in the arrays they themselves allocated in \texttt{dynk\_parseFUN}, i.e.\ the functions should not interact outside of calling each other.

\subsection{Initializing the functions: \texttt{dynk\_parseFUN}}
\label{sec:FUN:parse}

This function is responsible for parsing the user input (\texttt{FUN} statements), filling the \texttt{funcs\_dynk} table, and allocating memory in the \texttt{\{c|f|i\}expr\_dynk} arrays.
It is one of the most complex parts of the DYNK module, and must be modified whenever one wants to add new DYNK functions.
In \texttt{dynk\_parseFUN}, each function type is initialized by a block of code.
The selection of the code to execute depends on the function type listed in the second word of the relevant \texttt{FUN} statement.
Note that all DYNK functions first call two support subroutines \texttt{dynk\_checkargs} and \texttt{dynk\_checkspace}, in order to check that the number of arguments is as expected for this function, and that there is enough space in the \texttt{\{c|f|i\}expr\_dynk} arrays.
If there is not enough space, it is automatically allocated, as described in Section~\ref{sec:data:FUN}.

Please also note that the initialization of the \texttt{GET} function, which returns the initial setting of an element, is only fully initialized after the \texttt{dynk\_pretrack} subroutine has run.

\section{Support for collimation version}

The collimation version of SixTrack~\cite{colltrack} is able to work around the 64 particle limit by tracking multiple samples of particles, i.e.\ loading the first 64 particles and then calling \texttt{thin6d} to track them for all turns, then loading the next 64 particles etc., as illustrated in Figure~\ref{fig:thin6d}.
For this to work correctly, DYNK must reset all element attributes at the beginning of each tracking simulation.
This is accomplished in \texttt{dynk\_apply}, which at the first turn of the first sample saves the original values (retrieved using \texttt{dynk\_getvalue}), and then on the first turn of consecutive samples resets all elements and attributes touched by DYNK to the stored value.
Furthermore, the outputfile \texttt{dynksets.dat} is only written while processing the first sample.

\section{Checkpointing and restarting}
The DYNK module supports checkpointing and restarting.
For this to work, it must be able to truncate the \texttt{dynksets.dat} file to its position at the loaded restart point, and to store the current state of the DYNK functions.
As all the current states of the functions are stored in the \texttt{\{c|f|i\}expr\_dynk} arrays and in \texttt{fsets\_dynk}, these arrays are written by the checkpoint/restart routines.
Additionally, the array usage counters \texttt{n\{c|f|i\}expr\_dynk} are also saved and restored.
The rest of the state of DYNK is unchanged after initialization, and is thus recreated when reading the input files and initializing the simulations.

\section{Summary and outlook}

The DYNK module in SixTrack has been a success and has been in use for more than three years, enabling several new studies.
Its overall architecture is designed to be extensible, to make it easy to add new functions, and to support acting on new types of elements and element attributes.
The implementation of DYNK has also been taken as an opportunity to clean up several parts of the SixTrack code, as was done by implementing \texttt{initialize\_element}.
Furthermore, it not only supports the standard SixTrack, but also the collimation version and checkpoint/restart.

For the future, it is likely that several more functions and elements will be added, such as a wider variety of random distributions, and support for the beam-beam and multipole element.
It may be interesting to support the setting of a global non-integer turn shift, in order to easily study the effects on different bunches along the ring.
Furthermore, the performance of \texttt{dynk\_setvalue} and \texttt{dynk\_getvalue} might be improved by caching the element index, avoiding the search over all elements every time these functions are called.
Finally, since DYNK can now change the reference energy using the \texttt{GLOBAL-VARS} mechanism, the large subroutines \texttt{thin6dua} and \texttt{thck6dua} can now in principle be removed, reducing the amount of code duplication.

\end{document}